\documentclass[aps,twocolumn,nofootinbib,longbibliography]{revtex4-1}
\usepackage[babel]{csquotes}
\usepackage{graphicx}
\usepackage{amsmath,amssymb}
\usepackage[colorlinks,citecolor=blue,linkcolor=blue,urlcolor=blue]{hyperref}
\usepackage{mathrsfs}
\usepackage{enumerate}
\def\be{\begin{equation}}
\def\ee{\end{equation}}

\begin{document}

\title{Compact phase space, cosmological constant, discrete time}

\author{Carlo Rovelli${}^{a}$ and Francesca Vidotto${}^{b}$\vspace{.5em} }
\affiliation{\small
${}^a$\ \mbox{CPT, Aix-Marseille Universit\'e, Universit\'e de Toulon, CNRS,} \\
% \mbox{Samy Maroun Research Center for Time, Space and the Quantum,}  \\ 
Case 907, F-13288 Marseille, France. \vspace{.3em} \\
${}^b$\  \mbox{Radboud University, Institute for Mathematics, Astrophysics and Particle Physics,}
 \mbox{Mailbox 79, P.O. Box 9010, 6500 GL Nijmegen, The Netherlands.}}

\date{\small\today}

\begin{abstract}
\noindent  
We study the quantization of geometry in the presence of a cosmological constant, using a discretization with constant-curvature simplices.  Phase space turns out to be compact and the Hilbert space finite dimensional for each link. Not only the intrinsic, but also the extrinsic geometry turns out to be discrete, pointing to discreetness of time, in addition to space.  We work in 2+1 dimensions, but  these results may be relevant also for the physical 3+1 case.
\end{abstract}

\maketitle

\section{Introduction}

The presence of the cosmological constant can affect the quantum kinematics of gravity.  Here we show that it enters naturally in loop quantum gravity (LQG) by determining the size of a compact phase space and  the dimension of the corresponding finite dimensional Hilbert space. This yields the discretization of the extrinsic curvature and can be related to time discreetness. 

Recent results indicate that a positive cosmological constant simplifies, rather than complicating, our understanding of quantum gravity.  Fairbairn and Meusburger \cite{Fairbairn:2010cp} and Han \cite{Han:2010pz,Han:2011aa,Han:2013tap}, building on \cite{Rovelli1993,Smolin95} and \cite{Noui:2002ag}, have shown that  the cosmological constant makes covariant LQG finite.  Haggard, Han, Kami\'nski and Riello \cite{Haggard} have given a straightforward construction of the LQG  dynamics in the presence of the cosmological constant, related to the geometry of constant curvature triangulations, a key idea introduced by Bahr and Dittrich \cite{Bahr:2009qd}, which grounds the present work. The LQG kinematics needs to be modified to take into account the presence of a cosmological constant; this was realised long ago by Borissov, Major and Smolin \cite{Borissov:1995cn,Major:1995yz,Major:2007nd} and the problem has been recently explored in detail by Dupius, Girelli, Livine and Bonzom \cite{Dupuis2013,Dupuis2014,Bonzom2014,Bonzom2014a} for negative cosmological constant.

A discretization of spacetime in terms of flat simplices is not suitable for a theory with cosmological constant because flat geometry solves the field equations only with vanishing cosmological constant. This problem can be solved choosing a discretization with simplices with constant curvature.  Here we show with a positive cosmological constant, a constant curvature discretization leads to a modification of the LQG phase space. The phase space turns out to be \emph{compact} for each link. The  conventional LQG phase space is modified by curving the conjugate momentum space.  Curved momentum space has been repeatedly considered in quantum gravity, for instance in the  relative locality framework \cite{AmelinoCamelia:2011bm}. Here it is not the momentum space of a particle to be curved, but rather the space of the conjugate momentum of the gravitational field itself.  We study the quantization of the resulting phase space, and we write explicitly modified quantum geometrical operators. We show that these are related to a $q$ deformation of $SU(2)$.  A $q$ deformations has been derived in LQG as a way to implement the dynamics of the theory with cosmological constant in \cite{Noui2011,Pranzetti2014}.  Here we have shown that it is also directly implied by the constant curvature of the individual simplices, and we have given the corresponding form of the geometrical operators of the gravitational theory, in the presence of a cosmological constant.  

This result has physical consequences: it is not just the intrinsic geometry, but also the extrinsic geometry to be discrete in quantum gravity (for a hint of this see \cite{Rovelli2013d}).  This indicates that not just space, but also proper time, is predicted to be discrete in the theory.  

We study the quantization of the geometry in 2+1 Euclidean dimensions with positive cosmological constant, that results from building on constant curvature triangulations, but the results that we obtain may be relevant for the physical   3+1 Lorentzian case. 

\section{Constant curvature geometry}

We start with the geometry of a constant curvature triangle.  For this, it is convenient to fix units where the constant curvature has unit value. The small-curvature limit is then the limit where the triangle is small.  Consider therefore a metric sphere with unit radius $R=1$. A constant curvature triangle has three vertices on this sphere.  It is a portion of the sphere bounded by three arcs of maximal circles joining two vertices.  The geometry of the triangle is determined, up to isometries, by giving the length $L_l, l=1,2,3$, of these three (oriented) arcs.   Importantly, these lengths are bounded.  

Since the radius of the sphere determines a unit of length, these lengths can be given a-dimensionally: they are determined by the three angles $\alpha_l=L_l/R$ they define at the center of the sphere. There is also an elements $k_l$ of $SO(3)$ associated which each arc. This is simply the rotation by an angle $\alpha_l$ around the axis normal to the plane of the great circle of the arc. The intrinsic geometry of the triangle can therefore be given by associating (non independent) $SO(3)$ elements $k_l$ to each of its sides. 

Consider a two-dimensional surface $\Sigma$ immersed in a three dimensional Riemannian manifold. This surface inherits an intrinsic and an extrinsic geometry from the 3d manifold.  Fix a 3d triangulation of the 3d manifold, inducing a 2d triangulation on $\Sigma$. The holonomy  of the 3d spin connection along each side $l$ of the triangulation is an element $h_l$ of $SO(3)$. The index $l$ runs over all the segments (arcs) of the triangulation.  We choose to approximate the intrinsic geometry of each 3-cell with a constant curvature metric, and therefore the geometry of each triangle with a triangle of uniform curvature. The the full intrinsic and extrinsic geometry of $\Sigma$ is defined by a couple of $SO(3)$ elements, $(k_l,h_l)$, associated to each arc: $h_l$ is the holonomy of the 3d connection, $k_l$ is the rotation associated to the curved arc $l$. 

It is customary in loop quantum gravity (LQG) to consider the trivalent graph dual to the triangulation. In two space dimension, each link $l$ of the graph corresponds to an arc $l$ of the triangulation. The geometrical data are therefore an element $(k_l,h_l)$, in $SO(3)\times SO(3)$ for each link $l$ of the graph.  

In the limit in which the triangles are small compared to the constant curvature scale, the group elements $k_l$ are near the identity and we can approximate them as $k_l=e^{\vec J_l\cdot \vec\tau}\sim 1\!\!1 + J_l= 1+ \vec J_l\cdot \vec\tau$, where  $\vec\tau$ is a basis in the $so(3)\sim su(2)$ algebra. In this limit, which we call $R\to\infty$, the geometry of the surface is approximated by an element of the group, $h_l$, and an element of the algebra, $\vec J_l$, for each arc $l$. These are the standard  geometrical data of LQG. Let us shift from $SO(3)$ to $SU(2)$ as is conventionally done in LQG.  We write the limit in the form   
\begin{eqnarray}
SU(2) \times SU(2)  
&\underset{R\to\infty}{\to}
& su(2)\times SU(2)=T^*SU(2) \nonumber\\\nonumber
(k,h) &\underset{R\to\infty}{\mapsto}
& (J,h)
\end{eqnarray}
Thus, what we do here is to modify the usual LQG kinematics by replacing the algebra with the group.  The role of $SU(2)\times SU(2)$ in Euclidean 3d gravity with positive cosmological constant has been pointed out by Meusburger and Schroers in \cite{Meusburger2003,Meusburger2008}, as the local isometry group, or as the gauge group of the Chern-Simon formulation of the theory.

\section{Compact phase spaces}

The key difference between the algebra and the group is that the second is compact.  This has  significative consequences in the quantum theory.  Their importance for finiteness has been pointed out in \cite{Nozari2014}.  These are the consequences we explore here.  

A compact phase space is the classical limit of a quantum system with a finite dimensional Hilbert space. This can be seen in many ways; the simplest is to notice that a compact phase space has a finite (Liouville) volume, and therefore can accommodate a finite number of Planck size cells, and therefore a finite number of orthogonal quantum states.  The familiar example of quantum systems with finite dimensional Hilbert space is given by angular momentum, for systems with fixed total angular momentum, where the quantum state space is the Hilbert space ${\cal H}_j$ that carries the spin-$j$ representation of $SU(2)$. 

In standard LQG, the kinematical data are given by an element of $\Gamma\equiv su(2)\times  SU(2)$ on each link. $\Gamma$ is the phase space of the theory, for each link. Since it is a  a cotangent space, it carries a natural symplectic structure.  The corresponding quantization defines the quantum theory of gravity in the loop representation. This is defined on the Hilbert space $L_2[SU(2)]$, where the group elements act multiplicatively and the algebra elements act as left invariant vector fields.   Here we want to modify this structure by replacing the algebra $su(2)$ with the group $SU(2)$.  The problem we address is therefore to determine the phase space structure of $SU(2)\times SU(2)$ and its quantization.   

As a preliminary exercise, we address this problem for the simplified case of $U(1)\times U(1)$.

\section{$\mathbf{U(1)\times U(1)}$}

Here we define and quantize the phase space $U(1)\times U(1)$. We write elements of this space as a couple $(h,k)$ of complex numbers with unit norms, with $(h=e^{i\alpha},k=e^{i\beta})$. Let us start by determining the phase space structure, namely writing the symplectic two-form.  We are guided to do so by the fact that in the limit in which the radius of one of the two circles can be considered large we want to recover the symplectic form of the cotangent space, which is
\be
      \omega=d\alpha \wedge d\beta.
\ee
This indicates immediately what we need:
\be
      \omega=-h^{-1}dh \wedge k^{-1} dk.
\ee
which locally is just the same as the previous one.  The corresponding Poisson brackets are easily computed: 
\be
\{k,h\}=hk. 
\ee
This defines the phase space.  Let us look for a corresponding quantum theory.  For this, we want a Hilbert space $\cal H$ and operators $h$ and $k$ with an operator algebra that reduces to the above Poisson algebra in the appropriate limit.  The problem is easy to solve. 

The Hilbert space is the \emph{finite dimensional} Hilbert space $\cal H$ with a discrete basis $|n\rangle$ where $n=1,...,N=\dim {\cal H}$, and the operators act as follow
\be
   k|n\rangle =e^{i\frac{2\pi}N n}|n\rangle
\ee
and
\be
   h|n\rangle = |n+1\rangle
\ee
cyclically (that is $ k|N\rangle = |1\rangle$). A straightforward calculation gives their commutator algebra
\be
[h,k]= \left(e^{i\frac{2\pi}N}-1\right)  hk. 
\ee
which gives a representation of the Poisson algebra for large $N$. In this limit, the Planck constant (determined by $[\hat a,\hat b]=i\hbar \widehat{\{a,b\}}$) is related to $N$ by
\be
\hbar = \frac{2\pi}N. 
\ee
To understand the physics, recall that we have a preferred length here: the constant curvature radius, which we have set to unit.  Therefore the dimensionless quantity $\frac{2\pi}N$ is actually the ratio between two dimensionfull quantities: in the physical theory it is the ratio of the Planck length scale to the cosmological constant scale. 

The physics of the quantum theory of this simple example is intriguing. Since the phase space is compact, the Hilbert space is finite dimensional and therefore \emph{both} $h$ and $k$ have discrete spectrum.   This is like having a particle on a circle, and therefore discrete momentum, but also the circle being actually discrete, and therefore discrete position.  Discreteness of the position gives a maximum momentum. So, all physical quantities are discrete, bounded and therefore completely finite. 
 
This is appealing.   The mathematics of the continuum is just a useful approximation to a physical reality which is always discrete and finite.  

\section{$\mathbf{SU(2)\times SU(2)}$}

Our task is now to repeat the previous exercise for $\Gamma =SU(2)\times SU(2)$. We use the notation
\be
        (k=e^J,h)\in SU(2)\times SU(2),
\ee
In the $R\to \infty$ limit where the length of the arc is small compared to $R$, we have $k\sim 1\!\!1+ J$.
The symplectic structure of $T^*SU(2)$ is defined by the symplectic form $\omega=d\theta$ where 
\be
\theta=Tr[Jh^{-1}dh].  \label{thetazero}
\ee
There are several possibilities for deforming this structure to have it well defined on $SU(2)\times SU(2)$. The simplest possibility is to take 
\be 
    \theta= Tr [k h^{-1} dh]
\ee 
which reduces to \eqref{thetazero} in the limit. The resulting symplectic form 
\begin{eqnarray} 
\omega= Tr [dk \wedge h^{-1} dh 
- k h^{-1}dh \wedge h^{-1} dh] 
\end{eqnarray}
is closed and invariant under $h\to \lambda h$ with $\lambda\in SU(2)$. (It is not invariant under  $k\to \lambda k$, but this transformation is not a symmetry: it transforms small triangles into large ones). But there are others and more interesting phase space structures that one can write on the group $SU(2)\times SU(2)$.  The different forms of compatibility between phase space structure and group structure are studied under the name of Poisson-Lie groups and quasi Poisson-Lie groups \cite{Marmo1993,Lu1990,Alekseev2000,Alekseev2002,Meusburger2008,Schroers2011}.   

Instead of trying to guess physically interesting symplectic (or Poisson) structures, we  go directly to the quantum theory and study the problem there. We can then recover a phase space structure form the classical limit of the quantum operator algebra.  After all, in the real world it is the quantum theory to have a classical limit, not the other way around.

What we are seeking is therefore a deformation of the standard LQG operator algebra giving a \emph{finite} Hilbert space. For this, let us start recalling the action of the operators $h$ and $k=e^{J}$ on the standard LQG representation. This is,   
\be\label{U}
  h\psi(U)=U\psi(U)
  \ee
  and
  \be
    J^i\psi(U)=L^i\psi(U).
\ee
where $L^i$ is the left invariant derivative operator on the group manifold.  Let us  transform this to the canonical basis of $L_2[SU(2)]\sim \oplus_{j=0}^\infty ({\cal H}_j\otimes {\cal H}_j)
$
\be\label{base}
\langle U | jmn\rangle=D^j_{mn}(U)
\ee
defined by the Wigner matrices $D^j_{mn}$.  The result is 
\be
  J^i|jmn\rangle =  \tau^{(j)}_{mk} |jkn\rangle.  \label{J}
\ee
where $\tau^{(j)}_{mk}$ are the $SU(2)$ generators in the $j$ representation, and
\be  \label{hold}
  h_{AB}|jmn\rangle=\left(\begin{array}{ccc}
\frac12  & j  & j'  \\
A  & m  & m'  
\end{array} 
\right)\!\left(\begin{array}{ccc}
\frac12  & j  & j'  \\
B  & n  & n'  
\end{array}
\right) |j'm'n'\rangle
\ee
where $A,B$ are the indices of the $SU(2)$ matrix, repeated indices are summed over and the matrices are the Wigner 3j symbols.  Equation  \eqref{hold} is obtained from \eqref{U} and \eqref{base} by noticing that the group elements are the same as their spin-$\frac12$ representation and using the standard decomposition of products of representations. 

The first of these two equations is analogous to the abelian case: the operator is defined on the finite dimensional Hilbert space formed by a single spin component. 

The second equation, however, requires the quantum state space to be infinite dimensional, because there are non-vanishing Wigner 3j symbols anytime $j'=j\pm\frac12$. Can this equation be modified to adapt it to a finite dimensional Hilbert space? The answer is well known: let us  \emph{define} the action of the $h$ operator in the constant curvature case to be 
\be 
  h_{AB}|jmn\rangle=\left(\begin{array}{ccc}
\frac12  & j  & j'  \\
A  & m  & m'  
\end{array}
\right)_{\!\!q}\!\!
\left(\begin{array}{ccc}
\frac12  & j  & j'  \\
B  & n  & n'  
\end{array}
\right)_{\!\!q} |j'm'n'\rangle     \label{h}
\ee
where we have replaced the Wigner 3j symbols with their $q$ deformation, with $q^r=-1$ for an integer $r$ \cite{Kauffman1994}.  These operators are now well defined on the finite dimensional Hilbert space 
\be
   {\cal H}= \oplus_{j=0}^{j_{max}} ({\cal H}_j\otimes {\cal H}_j)
\ee
where 
\be
j_{max}=\frac{r-2}2.
\ee
Equations  \eqref{J} and \eqref{h} define the quantum theory in the constant curvature case, and represent our main proposal for the kinematics of quantum gravity in the presence of a cosmological constant. (Compare also with \cite{Major2008}.)

The $h$ operators no longer commute.  This can be seen using the graphical notation. Writing the Wigner symbols as a trivalent node, the matrix elements of the $h$ operator read 
\be
(h_{AB})_{m'n'}^{mn}\ \ =\ \ \raisebox{-6.5mm}{\includegraphics[width=20mm]{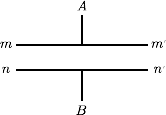}}~ .
\ee
If we act with two operators we have
\be
h_{AB}h_{CD}\ \ =\ ~ ~ ~ \raisebox{-6mm}{\includegraphics[width=15mm]{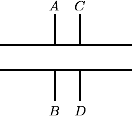}}~ .
\ee
while acting in the reverse order gives
\be
h_{CD}h_{AB}\ \ =\ ~ ~ ~ \raisebox{-6mm}{\includegraphics[width=15mm]{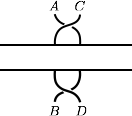}}~ .
\ee
In the $R\to\infty$ (or $q=1$) case, the crossing of two lines gives at most a sign, which squared gives unit. Therefore the two operators commute.  In the $q$ deformed case, the crossing gives a $q$ dependent factor  \cite{Kauffman1994,Roberts1995} and therefore the operators fail to commute.  If we call $R$ the operator giving the crossing, we have
\be
h_{AB}h_{CD}=R_{AC}^{A'C'} R_{BD}^{B'D'}\ h_{C'D'}h_{A'B'},
\ee
and if $R$ can be expanded in $\hbar$ as $R\sim 1+r$, we obtain classical Poisson brackets of the form 
\be
\{h_{AB},h_{CD}\}=r_{BD}^{B'D'} h_{CD'}h_{AB'}+r_{AC}^{A'C'} h_{C'D} h_{A'B}. 
\ee
Since the $h$ operators do not commute, there is no $h$ representation in the quantum theory anymore.

This completes the quantization of the $SU(2)\times SU(2)$ phase space. As in the abelian case, the resulting Hilbert space is finite dimensional. The dimension of the Hilbert space is determined by the ratio between the two constants: the one introduced by the quantization (physically: the Planck constant scale), and the constant curvature of the simplices, which enters via the deformation of the Poisson algebra -- physically: the cosmological constant. 

A $q$-deformation of the dynamics renders quantum gravity finite. This is known since the early nineties in 2+1 dimensions thanks to the Turaev-Viro state sum model  \cite{Turaev:1992hq}, which renders Ponzano Regge 2+1 quantum gravity finite, and whose strict connection to LQG was early pointed out \cite{Rovelli1993}.  The result has been extended to 3+1 dimensions in \cite{Fairbairn:2010cp,Han:2010pz,Han:2011aa,Han:2013tap}. The $q$-deformation of the dynamics amounts to the introduction of a cosmological constant \cite{Mizoguchi:1991hk,Smolin95}; in 2+1, the relation between the deformation parameter $q$ and the cosmological constant $\lambda$ is \cite{Mizoguchi:1991hk,Rovelli}
\be 
q=e^{i\sqrt{\Lambda}\hbar G}.
\ee
A $q$ deformations has been derived in LQG as a way to implement the dynamics of the theory with cosmological constant in \cite{Noui2011,Pranzetti2014}.  Here we have shown that a $q$-deformation is also directly implied by the constant curvature of the individual simplices, and we have given the corresponding form of the geometrical operators of the gravitational theory.  

\section{Physical considerations}

The Hilbert space $\cal H$ constructed above reduces to the usual LQG Hilbert space when the triangles are small compared to the curvature radius.  Namely when the region considered is small compared to the cosmological constant scale.   But there is a bound to this smallness, which is determined by the value of $j_{max}$, namely by the ratio of the cosmological constant scale to the Planck scale.   The region considered can never be smaller than the Planck scale and therefore never arbitrarily small, in the units in which the constant curvature radius is units.  (See  \cite{Bianchi:2011uq} for a discussion of the geometrical interpretation of this ratio.)

The effect of the compactness of $SU(2)$ in conventional LQG is the discretization of the intrinsic geometry.  In the quantization considered here, which takes the cosmological constant into account, there is a further compactness:  the entire phase space, and not just the configuration space, is compact.   Therefore also the variable conjugate to the intrinsic geometry is compact. This is evident form the fact that Hilbert space is finite dimensional (for each link), and therefore \emph{all} local operators have discrete spectrum.  Therefore the extrinsic geometry is quantized as well.  

The extrinsic curvature $K_{ab}$ determines the rate of change of the intrinsic geometry, because (in the Lapse=1, Shift=0 gauge) it is the proper-time derivative of the metric: $K_{ab}\sim dq_{ab}/dt$. Since $q_{ab}(\Delta t)\sim q_{ab}(0)+dq_{ab}/dt\ \Delta t$, we can infer the lapsed proper time $\Delta t$  from the values of $q_{ab}(0), q_{ab}(\Delta t)$ and $K_{ab}$. Since all these quantities have discrete spectrum, we expect proper-time intervals, measured using gravitational observables, to be discrete as well. While one expects the scale of minimum proper time to be Planckian, its full discrete spectrum can depend on the cosmological constant. This is similar to the angle discreteness pointed out by Seth Major \cite{Major2011a}.  

Do these results generalise to 3+1 dimensions? In the Euclidean case, the situation appears very similar. In the presence of a cosmological constant, we cannot choose a discretization of spacetime with flat simplices, because these are not solutions of the field equations. A constant curvature four simplex is bounded by constant curvature tetrahedra, and the geometry of these determines again a compact space, as for the curved triangles considered above. A compact space, in turn, determines a finite Hilbert space for each link. 

In the  Lorentzian case, the problem is more subtle, because of the hyperbolic geometry. However, it seems reasonable to require single cells of the triangulation to stay within the De Sitter horizon.  This again yields a maximal size, determined by the cosmological constant, and therefore a finite Hilbert space per each link. The situation, however, is still unclear in the Lorentzian 3+1 case. 

\vskip1mm
\centerline{$\sim\sim\sim$} 
\vskip1mm
{\em Note:} We understand that Han, Haggard, Kami\'nsky and Riello have related results in 3+1 dimensions, which will appear soon.

\vskip2mm
\noindent
{\bf Acknowledgements.}  ~ FV thanks Hal Haggard and Muxin Han for detailed explanations on \cite{Haggard}, before publication, which have inspired this works.  CR thanks Mait\'e Dupius, Florian and Ela Girelli, and  Aldo Riello, for extensive discussion on the topic. \\ FV acknowledges support from the Netherlands Organisation for Scientific Research (NWO) Veni Fellowship Program. CR acknowledges support from Samy Maroun Research Center for Time, Space and the Quantum.\\ 

\providecommand{\href}[2]{#2}\begingroup\raggedright\endgroup

%\bibliographystyle{utcaps}%{abbrv}\bibliographystyle{plain}
%\bibliography{library}

\end{document}